\documentstyle[12pt,amsbsy]{article}
\title{QUASI-EXACTLY SOLVABLE POTENTIALS WITH TWO KNOWN EIGENSTATES}
      
\author{V. M. Tkachuk \\
Ivan Franko Lviv State University, Chair of Theoretical Physics \\ 
		 12 Drahomanov Str., Lviv UA--290005, Ukraine\\
		 E-mail: tkachuk@ktf.franko.lviv.ua }

\begin{document}

\setlength{\jot}{1em}
\setlength{\abovedisplayskip}{1.5em}
\setlength{\belowdisplayskip}{1.5em}

\setcounter{page}{1}

\maketitle
\begin{abstract}
A new supersymmetry method for the generation of the quasi-exactly
solvable (QES) potentials with two known eigenstates is proposed.
Using this method we obtained new QES potentials for which
we found in explicit  form the energy levels and wave functions of the 
ground state and first excited state.

\end{abstract}

PACS number(s): 03.65.-w; 11.30.Pb

\section{Introduction}

Quasi-exactly solvable (QES) potentials for which a finite number
of eigenstates is explicitly known nowadays attract much attention.
This is an intermediate class between the problems for which the spectrum
can be found exactly and those which can not be solved.
The first examples of QES potentials were given in [1--4].
Subsequently several methods were worked out for generating QES potentials
and as a result many QES potentials were established [5--12].
One of the methods is the generation of new QES potentials using 
supersymmetric (SUSY) quantum mechanics [12--14]. This method applies
the technique of SUSY quantum mechanics (see review [15]) to calculate the
supersymmetric partner potential of the QES potential with $n+1$
known eigenstates. From the unbroken SUSY it follows that the
supersymmetric partner is a new QES potential with $n$ known
eigenstates. It is worth stressing that the starting point 
of the SUSY method used in [12--14]
for generating new QES potentials is the knowledge of the initial 
QES potentials. 
Note also that in \cite{14,14N} using SUSY quantum mechanics 
a various families of conditionally exactly solvable (CES) potentials
were constructed. 
The CES potentials are those for which the eigenvalues problem
for the corresponding Hamiltonian is exactly solvable only when the potential
parameters obey certain conditions \cite{15}.

In the present paper we propose a new SUSY technique for generating
QES potentials with the two known eigenstates. In contrast to the
previous paper [12--14] our SUSY method does not require the knowledge of
the initial QES potential for the generation of new QES one. As a result,
we obtained new QES potentials for which we found in
explicit form the energy levels and wave functions of the ground 
and first excited states.

\section{SUSY quantum mechanics and QES problems}

Let us first take a look at the Witten's model of SUSY quantum
mechanics. 
The SUSY partner Hamiltonians $H_\pm$ are given by
\begin{equation} \label{5}
H_\pm=B^\mp B^\pm=-{1\over2}{d^2\over dx^2}+ V_\pm(x),
\end{equation}
where
\begin{equation} \label{3}
B^\pm={1\over\sqrt{2}}\left(\mp{d\over dx}+ W(x)\right),
\end{equation}
$V_\pm(x)$ are the so-called SUSY partner potentials
\begin{equation} \label{6}
V_\pm (x)={1\over 2}\left(W^2(x) \pm W'(x)\right), \ \ W'(x)={dW(x)\over dx},
\end{equation}
$W(x)$ is the superpotential.

Consider the equation for the energy spectrum
\begin{equation} \label{7}
H_\pm \psi_n^\pm (x)=E_n^\pm \psi_n^\pm (x), \ \
n=0, 1, 2,... .
\end{equation}
The Hamiltonians $H_+$ and $H_-$ have the same energy spectrum
except the zero energy ground state which exists in the case 
of the unbroken SUSY. 
Only one of the Hamiltonians
$H_\pm$ has the zero energy eigenvalue. We shall use the convention
that the zero energy eigenstate belongs to $H_-$. 
The corresponding wave function 
due to the
factorization of the Hamiltonian $H_-$
satisfies the equation
$B^-\psi_0^-(x)=0$ and reads
\begin{equation} \label{9}
\psi_0^-(x)=C_0^-\ \exp\left(-\int W(x) dx\right),
\end{equation}  
$C_0^-$ is the normalization constant.
Here and below $C$ denotes
the normalization constant of the corresponding wave function.

From the normalization condition it follows that
\begin{equation} \label{10}
{\rm sign}(W(\pm \infty)) = \pm 1.
\end{equation}

The eigenvalue and eigenfunction of the Hamiltonians $H_+$ and $H_-$
are related by SUSY transformations
\begin{eqnarray}  
&& E_{n+1}^-=E_n^+,\ \ E_0^-=0,  \label{11} \\
&& \psi_{n+1}^-(x)={1\over \sqrt{E_n^+}}B^+\psi_n^+(x), \label{12} \\
&& \psi_n^+(x)={1\over \sqrt{E_{n+1}^-}}B^-\psi_{n+1}^-(x). \label{13}
\end{eqnarray}

The unbroken SUSY quantum mechanics, namely, 
the SUSY transformations 
are used for the exact calculation of the energy 
spectrum and wave functions (see review \cite{13}).
In the present paper we use SUSY quantum mechanics for 
the generation of the QES potentials
with the two known eigenstates. 

Suppose we study the Hamiltonian $H_-$,
whose ground state is given by (\ref{9}).
Let us consider the SUSY partner of $H_-$, i.e. the Hamiltonian $H_+$. 
If we calculate the ground state of $H_+$ we immediately find the first 
excited state of $H_-$ using  
the SUSY
transformations (\ref{11}), (\ref{12}), (\ref{13}). 
In order to calculate 
the ground state of $H_+$ let us rewrite it in the following form
\begin{equation} \label{14}
H_+=H_-^{(1)} + \epsilon =B^+_1 B^-_1+ \epsilon, 
\ \  \epsilon > 0,
\end {equation}
which leads to the following relation between potentials
energies
\begin{equation} \label{VV}
V_+(x)=V_-^{(1)}(x) + \epsilon,
\end{equation}
where 
$B_1^{\pm}$ are given by (\ref{3}) with superpotential $W_1(x)$,
similarly $V_-^{(1)}$ is given by (\ref{6}) with $W_1(x)$,
$ \epsilon$ is the energy of the ground state of $H_+$ 
since $H_-^{(1)}$ has zero energy ground state.

Using this procedure $N$ times we obtain $N$
excited energy levels and corresponding wave functions 
of $H_-$
in the following form
\begin{eqnarray} \label{En}
E^-_n =\sum^{n-1}_{i=0}\epsilon_i, \\ \label{psin}
\psi^-_n(x)=C^-_n B^+_0... B^+_{n-2}B^+_{n-1}
\exp \left(-\int W_n(x)dx\right),
\end{eqnarray}
where 
$n=1, 2,..., N$. 
In our notations $\epsilon_0 = \epsilon$,
$B_0^{\pm}=B^{\pm}$, $W_0(x)=W(x)$.
Operators $B_n^{\pm}$ are given by (\ref{3}) with the superpotentials
$W_n(x)$ which satisfy the set of equations
\begin{equation} \label{Wn}
W^2_n(x)+W'_n(x)=W^2_{n+1}(x)-W'_{n+1}(x) +2 \epsilon_n, \ \ 
n=0,1,...,N-1.
\end{equation}
This relations for superpotentials are the generalization
of equation (\ref{VV}) for the case of $N$ steps. 

The formulas (\ref{En}) and (\ref{psin}) can be considered 
as expressions for the energy levels and 
eigenfunctions for QES potential with $N$ energy levels.
But in order to calculate the eigenfunctions in explicit
form it is necessary to solve the set of equations 
for superpotentials (\ref{Wn}).
Unfortunately in general, it is not possible to determine the
superpotentials from (\ref{Wn}) for arbitrary $N$.

Previously the set of equations for $W_n(x)$ was solved
in the special cases of the so-called
shape invariant potentials \cite{18}
and self-similar potentials \cite{Shab,Spir}
and as a result many exactly solvable potentials
were obtained \cite{SS} (see also review \cite{13}).

We consider a more
general case and do not restrict ourselves to the 
shape invariant or self-similar potentials. 
A {\it novelty} of this paper is that we are obtaining
a general solution of (\ref{Wn}) for $N=1$.
Namely, the both superpotentials $W(x)$ and $W_1(x)$
in this case can be expressed via some function.
As a result it is possible to construct general expression for
QES potentials with explicitly known two eigenstates.
It is the subject of the next section.

\section{Constructing QES potentials with explicitly
known two eigenstates}
Let us consider set of equation (\ref{Wn}) for $N=1$.
In this case we have one equation for two superpotentials
$W(x)$ and $W_1(x)$
\begin{equation} \label{19}
W^2(x)+W'(x)=W^2_1(x)-W'_1(x) +2 \epsilon.
\end{equation}

Note, that (\ref{19}) is the Riccati equation 
which can not be solved exactly
with respect to $W(x)$ for a given $W_1(x)$
and vice versa. 
A new moment of this paper is that we can find 
such a pair of $W(x)$ and
$W_1(x)$ that satisfies equation (\ref{19}). For this purpose
let us rewrite equation (\ref{19}) in the following form
\begin{equation} \label{20}
W'_+(x)=W_-(x)W_+(x) +2\epsilon,
\end{equation}
where
\begin{eqnarray} \label{21}
W_+(x)=W_1(x) + W(x),\\ \nonumber
W_-(x)=W_1(x) - W(x).
\end{eqnarray}

This new equation can be easy solved with respect to $W_-(x)$
for a given $W_+(x)$ and vice versa.
In this paper we use the solution of equation (\ref{20}) with
respect to $W_-(x)$
\begin{equation} \label{22}
W_-(x)=(W'_+(x) - 2\epsilon )/W_+(x).
\end{equation}
Then from (\ref{21}) and (\ref{22}) we obtain the pair of
$W(x)$, $W_1(x)$ that satisfies equation (\ref{19})
\begin{eqnarray}  \label{23}
W(x)={1\over 2}\left(W_+(x) - (W'_+(x)-2\epsilon)/W_+(x) \right), \\ \nonumber
W_1(x)={1\over 2}\left(W_+(x) + (W'_+(x)-2\epsilon)/W_+(x) \right), 
\end{eqnarray}
here $W_+(x)$ is some function of $x$ for which the superpotentials
$W(x)$ and $W_1(x)$ given by (\ref{23}) satisfy
condition (\ref{10}).
As we see from (\ref{21}) $W_+(x)$ satisfies the same condition (\ref{10})
as $W(x)$ and $W_1(x)$ do.

Let us consider continuous function
$W_+(x)$. Because $W_+(x)$ satisfies 
condition (\ref{10}) the function $W_+(x)$ must have at least one zero. 
Then
as we see from (\ref{22}), (\ref{23})  $W_-(x)$, $W(x)$ and 
$W_1(x)$ have the poles. 
In order to construct the superpotential free of singularities
suppose that $W_+(x)$ has only one zero at $x=x_0$
with the following behaviour in the vicinity of $x_0$
\begin{equation} \label{24}
W_+(x)=W'_+(x_0)(x-x_0).
\end{equation}
In this case the pole of $W_-(x)$ and $W(x)$, $W_1(x)$ at $x=x_0$ can
be cancelled by choosing
\begin{equation} \label{25}
\epsilon = W'_+(x_0)/2.
\end{equation} 
Then the superpotentials free of singularities are
\begin{eqnarray}  \label{26}
W(x)={1\over 2}\left(W_+(x) - (W'_+(x)-W'_+(x_0))/W_+(x) \right), 
\\ \nonumber
W_1(x)={1\over 2}\left(W_+(x) + (W'_+(x)-W'_+(x_0))/W_+(x) \right). 
\end{eqnarray}

In the present paper we use this nonsingular solution for superpotentials
in order to obtain nonsingular QES potentials.
Substituting the obtained result for $W(x)$ (\ref{26}) 
into (\ref{6}) we obtain
QES potential $V_-(x)$ with explicitly known wave function of ground state
(\ref{9}) and wave function of the excited state. 
The latter can be calculated using (\ref{psin})
\begin{equation} \label{psi1}
\psi_1^-(x)=C^-_1\ W_+(x) \exp\left(-\int W_1(x) dx\right).
\end{equation}  
As we see from (\ref{psi1}) $\psi_1^-(x)$ has one node, 
because $W_+(x)$ has only one zero. Thus, $\psi_1^-(x)$ 
indeed is the wave function of first excited state. 

Note, that  
all expressions depend on the function $W_+(x)$. 
We may choose various functions $W_+(x)$ and obtain in a result
various QES potentials. 

To illustrate the above described method we give four explicit examples
of nonsingular QES potentials.
First example is the well known QES potential and is specially
chosen to show that our method works correctly.
In the next examples we present new QES potentials which as far
as we know have not been previously discussed in the literature.

\subsection{Example 1}

Let us first consider an explicit example which as we shall see
gives the well known QES potential. Let us put
\begin{equation}  \label{27}
W_+(x)=A\left({\rm sinh}(\alpha x)-{\rm sinh}(\alpha x_0) \right),
\end{equation}
where $A > 0$, $\alpha > 0$.
Then using (\ref{26}) we obtain
\begin{eqnarray}  \label{28}
W(x)={1\over 2}\left(A\left({\rm sinh}(\alpha x)-{\rm sinh}(\alpha x_0) 
\right)-
\alpha \ {\rm tanh}\left({\alpha \over 2} (x+x_0)\right) \right), \\ \nonumber
W_1(x)={1\over 2}\left(A\left({\rm sinh}(\alpha x)-{\rm sinh}(\alpha x_0) 
\right)+
\alpha \ {\rm tanh}\left({\alpha \over 2} (x+x_0)\right) \right),
\end{eqnarray}
which satisfy (\ref{10}).
Substituting this $W(x)$ into (\ref{6}) we obtain the potential 
energy 
\begin{equation}  \label{29}
V_-(x)={1\over 2}\left({1\over 4}A^2 \left({\rm sinh}(\alpha x)-
{\rm sinh}(\alpha x_0) \right)^2
-A \alpha {\rm cosh}(\alpha x)+
{1\over 2} A \alpha {\rm cosh}(\alpha x_0)+ 
{{\alpha}^2 \over 4} \right).
\end{equation} 
In the case of $\alpha > {1\over2}A$
this is a non-symmetric double-well potential, 
$x_0$ is responsible for an asymmetry of potential. 
In the case of $x_0=0$ we obtain the symmetric QES potential. 
It is the special case of Razavy potential \cite{3} with two known
eigenstates.
The QES potential (\ref{29}) was derived in \cite{16,17}
using the method elaborated in the quantum theory of spin systems 
(see also review \cite{9}). It is interesting to note, that as we show,
this potential has SUSY formulation.

We may calculate exactly the two eigenstates
for potential (\ref{29}). 
The distance between the ground energy level $E_0^-=0$ and the first
excited energy level $E_1^-$ is
\begin{equation}  \label{30}
\epsilon={1\over2}\alpha A {\rm cosh}(\alpha x_0).
\end{equation}
The wave function of the ground state can be easily calculated by (\ref{9})
\begin{equation} \label{31}
\psi_0^-(x) =C_0^- \ {\rm cosh} \left({\alpha \over 2}(x+x_0) \right)
\exp{\left(-
{A \over 2\alpha}{\rm cosh}(\alpha x) +
{A \over 2}{\rm sinh}(\alpha x_0)x \right)}.
\end{equation}
For the wave function of the first excited 
state using (\ref{psi1}) we obtain
\begin{equation} \label{32}
\psi_1^-(x) =C_1^- \ {\rm sinh} \left({\alpha \over 2}(x-x_0) \right)
\exp{\left(-
{A \over 2\alpha}{\rm cosh}(\alpha x) +
{A \over 2}{\rm sinh}(\alpha x_0)x \right)}.
\end{equation}

The results (\ref{30}), (\ref{31}), (\ref{32}) are the same as 
were obtained in \cite{17}
and we may claim that our SUSY method works correctly.

\subsection{Example 2}
Let us consider
\begin{equation} \label{Wp2} 
W_+(x)={A \ {\rm sinh}(\alpha x)\over b+c \ {\rm cosh}(\alpha x)},
\end{equation}
which
is a generalization of the function $W_+(x)$ considered 
in first example for $x_0=0$
and reproduces it in the special case $c=0$.
The parameters satisfy the condition
$A>0, \ \alpha >0, \ c>0$ and $b+c>0$. The last condition ensures
nonsingularity of $W_+(x)$.

The superpotentials $W(x)$ and $W_1(x)$ read
\begin{eqnarray}
W(x)={1\over 2}\left({(A+ac) \ {\rm sinh}(\alpha x)
\over b+c \ {\rm cosh}(\alpha x)} 
-{ab\over b+c} \ {\rm tanh}({\alpha x \over 2})\right), \\
W_1(x)={1\over 2}\left({(A-ac) \ {\rm sinh}(\alpha x)
\over b+c \ {\rm cosh}(\alpha x)} 
+{ab\over b+c} \ {\rm tanh}({\alpha x \over 2})\right).
\end{eqnarray}

The superpotential $W(x)$ generates the following QES potential
\begin{eqnarray} \label{V2}
V_-(x)={1\over 8c^2 }\left[
{(b^2-c^2)(A+ac)(A+3ac)\over (b+c \ {\rm cosh}(\alpha x))^2}
-{2b(A+ac)^2\over b+c \ {\rm cosh}(\alpha x)} \right. \\ \nonumber
\left.
+{a^2 b c^3\over (b+c)^2}{1\over {\rm cosh}^2(\alpha x/2)} 
+{(ac^2+A(b+c))^2 \over (b+c)^2}
\right].
\end{eqnarray}

The energy of the ground and first excited states are $E_0^-=0$ 
and $E_1^-=\epsilon =aA/2(b+c)$ respectively.

The wave function of the ground and first excited states read
\begin{eqnarray}
\psi_0^-(x)=C_0^-({\rm cosh}(\alpha x/2))^{b/(b+c)} 
(b+c \ {\rm cosh}(\alpha x))^{-1/2-A/2ac}, \\
\psi_1^-(x)=C_1^-{\rm sinh}(\alpha x)({\rm cosh}(\alpha x/2))^{-b/(b+c)} 
(b+c \ {\rm cosh}(\alpha x))^{-1/2-A/2ac}.
\end{eqnarray}

The wave functions of the ground state is square integrable for
any parameters of superpotential that satisfy the condition 
described above. The wave function of first excited
state is square integrable when
\begin{equation}
{c\over b+c}<{A\over ac}.
\end{equation} 
In the opposite case the system has only localized ground state.

As far as we know the potential in general form (\ref{V2})
has not been previously discussed in the literature.
This potential is interesting from that point of view that
in special cases of parameters it reproduces
the potentials studied early.
Thus, in the limit $b\to 0$ the superpotential $W_+(x)$ (\ref{Wp2})
generates the well known exactly solvable Rosen-Morse potential.
For $c\to 0$ one obtains the Razavy potential with two
explicitly known eigenstates (Example 1).

It is interesting to consider the special case $A=ac$. 
In this case the first term
in $W_1(x)$ drops up and $W_1(x)$ generates the SUSY partner
$V_+(x)$ which is Rosen-Morse potential and can be solved
exactly. Then using SUSY transformation (\ref{11}) and (\ref{12}) we
can easily calculate the energy levels and wave functions
of all states of Hamiltonian $H_-$ with potential energy $V_-(x)$.
In this special case $V_-(x)$ can be treated as CES potential
and it corresponds to the one studied in \cite{14N}.

\subsection{Example 3}
Consider the function $W_+(x)$ in the polynomial form
\begin{equation}
W_+(x)=ax + bx^3,
\end{equation}
where $a>0, b>0$.
The final result for QES potential is the following
\begin{equation}
V_-(x)={1\over 8}(a^2-12b)x^2+{ab\over 4}x^4+{b^2\over 8}x^6
+{3ab\over 8(a+bx^2)^2}+{3b\over 8(a+bx^2)}-{a\over 4}.
\end{equation}

The energy levels of the ground and first excited states are
$E_0^-=0$, $E_1^-=a/2$. Note, that two energy levels of this potential
do not depend on the parameter $b$. The wave functions of those
states read
\begin{eqnarray}
\psi _0^-(x)=C_0^- (a+bx^2)^{3/4}e^{-x^2(2a+bx^2)/8}, \\ 
\psi _1^-(x)=C_1^- x(a+bx^2)^{1/4}e^{-x^2(2a+bx^2)/8}.
\end{eqnarray}
It is worth to stress that the case $b=0$ corresponds to linear 
harmonic oscillator.

\subsection{Example 4}
Let us put
\begin{equation}
W_+(x)={Ax\over \sqrt{b^2+x^2}}, \ \ A>0.
\end{equation}
It is obviously that it is enough to consider only $b>0$.

The QES potential in this case reads
\begin{equation}
V_-(x)=
{1-A^2b^2\over 8(b^2+x^2)}- {Ab^2\over 2(b^2+x^2)^{3/2}}
-{5b^2\over 8(b^2+x^2)^2} + {(1+Ab)^2\over8b^2}.
\end{equation}

The energy levels of the ground and first excited states are
$E_0^-=0$, $E_1^-=a/2b$. The wave functions of those
states read
\begin{eqnarray}
\psi_0^-(x)=C_0^- \left(1+{b\over \sqrt{b^2+x^2}}\right)^{1/2}
e^{-\sqrt{b^2+x^2}(1+Ab)/2b}, \\
\psi_1^-(x)=C_1^- {x\over\sqrt{b^2+x^2}}
\left(1+{b\over \sqrt{b^2+x^2}}\right)^{-1/2}
e^{-\sqrt{b^2+x^2}(-1+Ab)/2b}.
\end{eqnarray}
The wave function of first excited state is square integrable
if $Ab>1$. Otherwise only the localized ground state
exists.

Thus as we see the proposed new SUSY method for constructing
QES potentials gives us an opportunity to obtain new potentials
for which two eigenstates are exactly known. In special cases our 
potentials reproduce those studied earlier.

\section{Concluding remarks}
In the present paper we propose a new SUSY method for constructing QES
potentials with two explicitly known eigenstates. Namely, we obtain
a general expression for QES potentials and wave functions of
the ground and first excited states. This method is more general
than those given in the literature before and in contrast to
them does not require the knowledge of the initial QES potentials
for constructing new QES potentials. But we must note that the
proposed method is restricted to constructing QES potentials with
explicitly known only two eigenstates.
Naturally, there is a question about the generalization of this
SUSY method for the case of QES potentials with explicitly known more
than two eigenstates. 
In the case of three energy levels from (\ref{Wn})
(where $N=2$ corresponds to two excited levels) we obtain
the set of two equations which relate three superpotentials.
In order to obtain QES potential with three known eigenstates
in explicit form we must solve this set of equations.
It is more complicated problem than the case of one
equation ($N=1$) which was considered in the present paper.
This problem will be a subject of a separate paper.

\pagebreak

\end{document}